\documentclass[prl,twocolumn,floatfix,superscriptaddress]{revtex4-1}
\usepackage{dcolumn,amsmath}
\usepackage{graphicx}
\usepackage{bm}
\usepackage{hyperref}

\begin{document}
\title{Scalar-pseudoscalar interaction in the francium atom}

\author{L.V.\ Skripnikov}\email{leonidos239@gmail.com}
\homepage{http://www.qchem.pnpi.spb.ru}
\author{D.E.\ Maison}
\affiliation{National Research Centre ``Kurchatov Institute'' B.P. Konstantinov Petersburg Nuclear Physics Institute, Gatchina, Leningrad District 188300, Russia}
\affiliation{Saint Petersburg State University, 7/9 Universitetskaya nab., St. Petersburg, 199034 Russia}
\author{N.S.\ Mosyagin}
\affiliation{National Research Centre ``Kurchatov Institute'' B.P. Konstantinov Petersburg Nuclear Physics Institute, Gatchina, Leningrad District 188300, Russia}

\date{28.11.2016}

\begin{abstract}
Fr atom can be successively used to search for the atomic permanent electric dipole moment (EDM) [Hyperfine Interactions 236, \textbf{53} (2015); Journal of Physics: Conference Series 691, 012017 (2016)]. 
It can be induced by the permanent electron EDM predicted by modern extensions of the standard model 
to be nonzero at the level accessible by the new generation of EDM experiments. 
We consider another mechanism of the atomic EDM generation in Fr. This is caused by the scalar-pseudoscalar nucleus-electron neutral current interaction with the dimensionless strength constant, $k_{T,P}$. Similar to the electron EDM this interaction violates both spatial parity and time-reversal symmetries and can also induce permanent atomic EDM. It was shown in [Phys. Rev. D \textbf{89}, 056006 (2014)] that the scalar-pseudoscalar contribution to the atomic EDM can dominate over the direct contribution from the electron EDM within the standard model.  We report high-accuracy combined all-electron and two-step relativistic coupled cluster treatment of the effect from the scalar-pseudoscalar interaction in the Fr atom. Up to the quadruple cluster amplitudes within the the coupled cluster method with single, double, triple, and noniterative quadruple amplitudes, CCSDT(Q), were included in correlation treatment.
 This calculation is required for the interpretation of the experimental data in terms of $k_{T,P}$. The resulted EDM of the Fr atom expressed in terms of $k_{T,P}$ is $d_{\rm Fr}= k_{T,P} \cdot 4.50\cdot10^{-18} e\cdot {\rm cm}$, where $e$ is the charge of the electron.
 The value of the ionization potential of the $^2S_{1/2}$ ground state of Fr calculated within the same methods is in very good agreement with the experimental datum.
\end{abstract}
%

\maketitle

\section{Introduction}

The Fr atom is considered as a potential system to search for the electron electric dipole moment (EDM) and the corresponding magneto-optical trap experiment is under preparation in Japan \cite{Kawamura:14,Kawamura2015,Harada:2016}. The nonzero value of the electron EDM implies manifestation of effects which violate both time-reversal (T) and spatial parity (P) symmetries.
Such a search is one of the key tests of the standard model and its popular extensions \cite{Commins:98,Chupp:15}. It was estimated in \cite{Sandars:64, Sandars:65,Khriplovich:11,Flambaum:76} that electron EDM is significantly enhanced in heavy atoms (with open shell and nonzero total moment). The enhancement factor scales roughly as $Z^3$ \cite{Flambaum:76} in the main group atoms where $Z$ is the nucleus charge. Therefore, the Fr atom being the heaviest available alkali-metal atom is one of the best possible candidates for such a search.

The feature of all experiments to search for the permanent EDM (both atomic and molecular) is that the enhancement factor which is required for the interpretation of the experimental data 
(value of permanent atomic EDM) in terms of electron EDM
cannot be obtained from direct experimental measurement and can be only calculated.
Such calculations of the enhancement factor have been performed for Fr ($Z=87$) in a number of papers \cite{Mukherjee:09,Roberts:13,Byrnes:99,Blundell:2012} where this factor was found to be of order of
$10^3$ which is about two times larger than that in the Tl atom ($Z=81$) \cite{Liu:92}.

We consider another source of the permanent EDM of Fr. This is the scalar-pseudoscalar nucleus-electron neutral current interaction which does not depend on the nucleus spin.
As was shown in Ref.~\cite{Pospelov:14} the contribution from this interaction far exceeds that 
from the electron EDM within the standard model.
We have performed benchmark relativistic correlation calculations of the electronic structure of Fr and found the conversion factor that is required to interpret the experimental data in terms of the characteristic dimensionless constant $k_{T,P}$ of the considered interaction.
Besides, the value of the ionization potential is calculated at the same level of accounting for the correlation and relativistic effects.

\section{Theory}
The T,P-odd scalar-pseudoscalar nucleus-electron interaction is given by the following operator (see Eq.~(130) in~\cite{Ginges:04}):
\begin{eqnarray}
 \hat{H}_{T,P}=k_{T,P}\cdot i\frac{G_F}{\sqrt{2}}Z \sum_p\gamma^0_p\gamma^5_p\rho_N(\textbf{r}_p) = k_{T,P}\cdot \hat{h}_{T,P},
 \label{Htp}
\end{eqnarray}
where $p$ is an index over electrons, $G_F$ is the Fermi-coupling constant, $\gamma^0$ and $\gamma^5$ are the Dirac matrices, and $\rho_N(\textbf{r})$ is the nuclear density normalized to unity. 
This interaction induces the atomic EDM, $d_{\rm{Atom}}$, and leads to the linear Stark shift, $\Delta E$, in a weak uniform external electric field directed by axis $z$ with magnitude $E_{\rm{ext}}$:
\begin{eqnarray}
\Delta E = E_{\rm{ext}} \cdot d_{\rm{Atom}}.
 \label{AtomEDM}
\end{eqnarray}
Here
\begin{eqnarray}
 d_{\rm{Atom}}=  
\sum_{j>0}
\frac{\langle \Psi_0 | \sum_p e\hat{z_p} | \Psi_j \rangle \langle \Psi_j |\hat{H}_{T,P} | \Psi_0
 \rangle}{E_0 - E_j}+h.c.\\
  =k_{T,P} \cdot R_s,
 \label{AtomEDM2}
\end{eqnarray}
where 
\begin{eqnarray}
 R_s= \sum_{j>0} \frac{\langle \Psi_0 | \sum_p e\hat{z_p} | \Psi_j \rangle \langle \Psi_j |\hat{h}_{T,P} | \Psi_0 \rangle}{E_0 - E_j}+h.c.
 \label{Rs}
\end{eqnarray}
is the constant that is required to interpret the experimentally measured atomic EDM 
in terms of the $k_{T,P}$ constant, h.c.\ means Hermitian conjugate, $e$ is the (negative) charge of the electron, $\hat{z}$ is the $z$-component of the dipole moment operator, $\Psi_0$ is the exact (correlated) many-electron wavefunction of the considered electronic state ($^2S_{1/2}$) of the Fr atom, $\Psi_j$ are the excited state wavefunctions of Fr, $E_0$ and $E_j$ are the corresponding exact total energies.

The direct use of Eq. (\ref{Rs}) corresponds to the so-called sum over states method.
Formally, the summation in this equation should be done for all the excited states.
Unfortunately, only the main contributions to this sum (see, e.g.,\ \cite{Byrnes:99}) 
are taken into account in practice.

We use an alternative approach and rewrite $R_s$ in the following form:
\begin{eqnarray}
 R_s=\frac{\partial^2 \Delta E}{\partial E_{\rm{ext}} \partial k_{T,P}}(E_{\rm{ext}}=0).
 \label{RsDer}
\end{eqnarray}
Thus, $R_s$ can be computed numerically as the mixed derivative.
Note that experimenters also use the ``finite difference'' technique as the typical value of the
external electric field is 100 kV/cm$\approx 1.94\cdot10^{-2}$ a.u. \cite{Regan:02}. 

The scalar-pseudoscalar interaction (as well as the electron EDM) leads to a nonzero effect in the first order only for the open-shell systems (molecules or atoms). One can see from Eq.(\ref{Htp}) that the matrix elements of $\hat{h}_{T,P}$ are determined by the behavior of the valence wavefunction mainly at the heavy nucleus. We call the corresponding properties as atoms-in-compounds (AIC) ones \cite{Skripnikov:15b,Titov:14a,Zaitsevskii:16a}. In Ref.~\cite{Skripnikov:16b} we have proposed the combined scheme of calculating such parameters. This scheme implies application of both the four-component Dirac-Coulomb(-Gaunt) and two-step approaches. The former is used to obtain the leading effects while the latter allows us to account approximately for high-order correlation effects and other corrections.
In the first stage of the two-step approach one uses the generalized relativistic effective core potential (GRECP) method  \cite{Titov:99,Mosyagin:10a,Mosyagin:16} to obtain the most accurate approximation of the wavefunction in the valence and outercore region. 
This is achieved by the following main features of the GRECP method:
(i) exclusion of the inner-core electrons of the considered heavy atom to reduce the number of correlated electrons. 
(ii) The calculations with the GRECP are two-component.
Therefore, one can use considerably smaller basis sets with respect to the 4-component approach where the additional basis set for the small components have to be used.
Note, however, that formally, the computational expenses of the correlation stage at the level of such methods as the coupled cluster with single, double and perturbative triple amplitudes are higher than that of the four-index transformation stage \cite{Fleig:12} in molecular programs. In practice, the computation time of each of the stages will also depend on the efficiency of parallelization of the four-index transformation and correlation calculation as well as parameters of computer hardware. As a result in some \textit{practical} cases the four-index transformation stage can be even more wall-clock time-consuming than the correlation stage.
(iii) Valence one-electron wavefunctions (spinors) are smoothed in the inner-core region. This allows us to reduce further the size of the basis set.
(iv) One can easily omit the spin-orbit part of the GRECP operator and use the scalar-relativistic approach, e.g., to generate the natural basis set \cite{Skripnikov:13a}, estimate different correlation contributions, choose the most appropriate method to account for electron correlation \cite{Skripnikov:16b}, etc. 
One should also note that compact \textit{contracted} basis sets can be used in GRECP calculations 
in practice. This is not always available in some four-component (molecular) codes.
Valence properties can be directly obtained from the GRECP calculations
\cite{Petrov:14,Skripnikov:15d,Cossel:12,Skripnikov:09a}.

It was noted above that the AIC properties are mainly determined by the valence wave function in the inner-core spatial region of the heavy atom. Therefore, after the first stage (GRECP calculation) one should restore correct four-component behavior of the valence wave function in the core region of the heavy atom. For this one can use the nonvariational restoration procedure developed in \cite{Titov:06amin,Skripnikov:15b,Skripnikov:16a,Skripnikov:11a}. The procedure is based on the approximate
proportionality of the valence and low-lying virtual spinors in the inner-core region of the heavy atom \cite{Titov:06amin}. The procedure has been recently extended to three-dimensional periodic structures (crystals) in Ref.~\cite{Skripnikov:16a}; it was also successfully used for precise investigation of different diatomics~\cite{Lee:13a,Skripnikov:15b,Skripnikov:14c,Petrov:13,Kudashov:13,Kudashov:14,Skripnikov:09,Skripnikov:15c,Skripnikov:14a,Skripnikov:13c,Le:13,Skripnikov:08a}.

The two-step method allows one to consider high-order correlation effects and very flexible
valence basis sets with rather modest requirements to computer resources in comparison to four-component approaches. However, some uncertainty remains due to the impossibility to consider the full version of the GRECP operator in the codes used in the present paper and neglect of the inner-core correlation effects. In Ref.~\cite{Skripnikov:16b} we suggested combining the two-step approach and the direct relativistic Dirac-Coulomb(-Gaunt) approach to take advantages of both approaches.

In Refs.~\cite{Skripnikov:16b,Skripnikov:15a} we analyzed different approaches to treat electron correlation effects for the problem of calculating the AIC characteristics.
We showed that the coupled cluster series gives very accurate results and converges fast with increasing of the order of the highest included cluster amplitudes.
Therefore, the single-reference coupled cluster approaches were also used in the present paper.
We do not use such approximations as linearization, etc., in the coupled cluster study.

The {\sc dirac12} code \cite{DIRAC12} was used for the Dirac-Fock calculations and integral transformations.
Relativistic correlation calculations were performed within the {\sc mrcc} code  \cite{MRCC2013}. For scalar-relativistic calculations we used the {\sc cfour} code \cite{CFOUR,Gauss:91,Gauss:93,Stanton:97}. 
The three GRECP versions with 78, 60, and 28 
($1s^2 2s^2 2p^6 3s^2 3p^6 3d^{10} 4s^2 4p^6 4d^{10} 4f^{14} 5s^2 5p^6 5d^{10}$,
$1s^2 2s^2 2p^6 3s^2 3p^6 3d^{10} 4s^2 4p^6 4d^{10} 4f^{14}$ and $1s^2 2s^2 2p^6 3s^2 3p^6 3d^{10}$, respectively) electrons in the inner cores (excluded from the following GRECP calculations) of Fr 
have been generated in the present paper following the technique \cite{Titov:99, Mosyagin:16}.
The principal distinctive features of the GRECP technique with respect to other RECPs are {\it (a)} 
generation of the potentials for both the valence and outercore electrons with different $n$ but with the same $lj$ pair and {\it (b)} addition of nonlocal (separable) terms to the conventional semilocal RECP operator. In particular, the potentials describing the states of the $5s$, $6s$, $7s$, $5p$, $6p$, $7p$, $5d$, $6d$, $5f$, and $5g$ electrons 
were constructed in the framework of the GRECP with the 60-electron inner core.
Then, the valence GRECP version was derived from the full GRECP one 
by neglecting the differences of the outercore potentials 
from the valence ones. Thus, the valence GRECP 
operator is the semilocal one with the $7s$, $7p$, $6d$, $5f$, and $5g$ 
components of the full GRECP version.
The main difference of the valence GRECP from the conventional RECPs is that the components 
of the former were constructed for nodal valence pseudospinors. Thus, 
these are the valence potentials (not the outercore or somehow averaged 
ones) which act on the valence electrons in this GRECP version. 
The contributions of the Breit interactions and Fermi nuclear charge 
distribution are approximately taken into account with the help of 
the GRECP operator. It was demonstrated \cite{Mosyagin:00, Isaev:00, Mosyagin:10c, Skripnikov:13a} 
that the GRECP method allows one to reproduce the results of the corresponding all-electron calculations with good accuracy (at significantly smaller computational costs).

The code to compute one center matrix elements of the scalar-pseudoscalar Hamiltonian over the atomic four-component spinors (bispinors) has been developed in the present paper.

\section{Calculation}

Our combined scheme included the following steps to calculate the $R_s$ constant that is required to interpret permanent atomic EDM caused by the scalar-pseudoscalar interaction in terms of the strength constant $k_{T,P}$.
(i) The main contribution has been calculated within the coupled cluster method with single, double and perturbative triple amplitudes, 
CCSD(T) with the use of the all-(87)-electron Dirac-Coulomb Hamiltonian.
The CVTZ \cite{Dyall:07,Dyall:12} basis set which consists of [34$s$,30$p$,19$d$,12$f$,1$g$] basis (uncontracted Gaussians) functions was used. The cutoff for the virtual atomic bispinor energies equal to 6500 a.u. was applied.
(ii) To consider the contribution to $R_s$ from the extension of the basis set we have frozen 28 electrons ($1s^22s^22p^63s^23p^63d^{10}$) of Fr and performed the
59-electron CCSD(T) calculations with the cutoff set to 100 a.u. (energy of 4$s$ atomic orbital is about -43 a.u.) within the Dirac-Coulomb approach. The considered contribution was obtained as the difference between the $R_s$ values calculated using the AE4Z and CVTZ basis sets. The AE4Z basis set \cite{Dyall:07,Dyall:12} contains [37$s$,35$p$,23$d$,16$f$,10$g$,4$h$,1$i$] basis functions.
(iii) The contribution from increasing the virtual spinor energy cutoff from 100 a.u. to 153 a.u. was calculated at the CCSD level and the AE4Z basis set  within the Dirac-Coulomb approach.
(iv) Influence of the Gaunt interaction on $R_s$ has been estimated at the self-consistent field level.
(v) The contribution to $R_s$ from high-order correlation effects was calculated as the difference in the calculated $R_s$ values at the 27-electron coupled cluster method with single, double, triple, and noniterative quadruple amplitudes, CCSDT(Q), versus the 27e-CCSD(T) method within the two-step approach.
We used the natural CBas basis set consisting of (25,25,20,15,10)/[10$s$,7$p$,4$d$,3$f$,1$g$] contracted functions (the numbers in the round and squared brackets refer to primitive and contracted functions, respectively), which has been generated using the technique developed in Ref. \cite{Skripnikov:13a}.
(vi) Contributions to $R_s$ from additional (with respect to those in the AE4Z basis set) $f$, $h$, and $i$ basis functions. The correction was calculated using the 59-electron scalar-relativistic CCSD(T) method within the two-step approach. 
Two basis sets were generated: LBas and LBasExt. The former consists of [25$s$,25$p$,20$d$,16$f$,10$g$,4$h$,1$i$] basis functions, where $f$,$g$,$h$ and $i$ basis functions are taken from the AE4Z basis set.
The LBasExt [25$s$,25$p$,20$d$,21$f$,10$g$,7$h$,6$i$] basis set was obtained from the LBas one by addition of 5f, 3h and 5i functions.
  
  In all the calculations atomic spinors were obtained using the average-of-configuration Dirac-Fock method for the one electron in the two spinors (one Kramers pair) (this is equivalent to the restricted Dirac-Fock approach), i.e. Dirac-Fock for the ground $^2S_{1/2}$ (7s) term of Fr.

Thus, the total of parameter $X$ ($X=R_s$ or IP) of the ground state of Fr were calculated using the following expression:
\begin{equation}
\begin{array}{l}
X({\rm TOTAL}) = X(\mbox{87e-4c-CCSD(T), CVTZ,6500 a.u.})  \\
                  \\   
                 + X(\mbox{59e-4c-CCSD(T), AE4Z, 100 a.u.}) -\\
                  X(\mbox{59e-4c-CCSD(T), CVTZ, 100 a.u.}) \\
                  \\
                 + X(\mbox{59e-4c-CCSD, AE4Z, 153 a.u.}) - \\
                 X(\mbox{59e-4c-CCSD, AE4Z, 100 a.u.}) \\
                 \\
                 + X(\mbox{4c-Dirac-Fock-Gaunt, AE4Z}) - \\
                  X(\mbox{4c-Dirac-Fock, AE4Z}) \\
                  \\
                 + X(\mbox{two-step-2c-27e-CCSDT(Q),CBas}) - \\ 
                 X(\mbox{two-step-2c-27e-CCSD(T),CBas}) \\
\\
                 + X(\mbox{two-step-1c-59e-CCSD(T),LBasExt}) - \\
                  X(\mbox{two-step-1c-59e-CCSD(T),LBas}) \\
\end{array}
 \label{Calc}
\end{equation}
The quantum electrodynamical (QED) self-energy and vacuum polarization contributions obtained in \cite{Shabaev:13,Tupitsyn:16} within the CI+MBPT method using the model QED potential approach were also taken into account for IP. 


\section{Results and discussion}

Table \ref{IPResults} presents the obtained results for the ionization potential in comparison with the previous calculations. 
\begin{table}[!h]
\caption{Values of ionization potential (IP) obtained within different approaches. IP(QM) is the IP obtained within relativistic quantum mechanics and the IP(QM+QED) is the IP corrected by QED contribution ($-5.07$ meV) from Refs.~\cite{Shabaev:13,Tupitsyn:16} (in meV).}
\label{IPResults}
\begin{tabular}{lll}
	\hline\hline
Method     & IP(QM)  & IP(QM+QED) \\
	\hline
	 Correlation potential \cite{Dinh:08a} & 4082.9 & 4077.9 \\ 
	 FS-CCSD \cite{Kaldor:1994}           & 4071.5 & 4066.4 \\ 
     This work, CCSD(T)       &        &        \\
     ~~~~~~~~~~~~~      +corrections       & 4071.2 & 4066.1 \\
	 Experiment \cite{Sansonetti:07}        &        & 4072.7\\
	\hline \hline
\end{tabular}
\\
\end{table}
One can see good agreement with the experimental data. Note that the IP in our approach was derived as the difference between the total energies obtained in different 
unrestricted coupled cluster calculations of the neutral and ionized Fr, i.e., all cluster amplitudes were optimized independently. 
%
In Table \ref{IPResults} we included the IP values corrected by the QED (self-energy and vacuum polarization) contributions obtained in Refs.~\cite{Shabaev:13,Tupitsyn:16}. 
The QED corrections obtained in Refs.~\cite{Dinh:08,Thierfelder:2010} are of the same order of magnitude.

The second derivative in Eq.~(\ref{RsDer}) can be calculated numerically within the following two strategies with respect to the method of including the interaction of the Fr electrons with the external electric field.
In the strategy (I) one adds the field at the self-consistent field stage while in the strategy (II) the field is added after the self-consistent field calculation. In the former case one considers the ``relaxation'' effects from the very beginning. Table \ref{TComparison} presents comparison of some correlation contributions obtained within the two approaches.
\begin{table}[!h]
\caption{Correlation contributions to $R_s$ (in $10^{-18} e \cdot {\rm cm}$) within the 27-electron coupled cluster methods using the two strategies of including the interaction with the external electric field (see the main text for details).
}
\label{TComparison}
\begin{tabular}{lrr}
\hline\hline
Contribution                            & Strategy I & Strategy II   \\
\hline
PT2$^{a,b}$                                    & 5.21    &   1.17   \\
\\
CCSD$^{b}$                                    & 4.34    &   4.23   \\
CCSD(T) - CCSD$^{b}$                            & 0.00    &  -0.10   \\
CCSDT - CCSD(T)$^{c}$                           &-0.08    &   0.12   \\
CCSDT(Q) - CCSDT$^{c}$                          & 0.00    &   0.00   \\
Sum                                     & 4.25    &   4.25   \\
\hline\hline
\end{tabular}
\\$^{a}$ Estimated as the first iteration of the CCSD calculation.
\\$^{b}$ Calculated within the four-component approach using the CVTZ basis set.
\\$^{c}$ Calculated within the two-step two-component approach using the CBas basis set.
\end{table}
 It follows from the table that the two strategies give almost equal results if high-order correlation effects are considered. However, the strategy (I) leads to the smoother convergence, therefore the main calculations below were performed within this approach.

Table~\ref{TResults} presents the obtained contributions to $R_s$ and IP following Eq.(\ref{Calc}).
\begin{table}[!h]
\caption{Values and contributions to $R_s$ and ionization potential of the ground state of Fr.}
\label{TResults}
\begin{tabular}{lll}
\hline\hline
Contribution                             & $R_s$, $10^{-18} e \cdot {\rm cm}$ & IP, meV \\
\hline
87e-4c-CCSD(T),   CVTZ                   & 4.51   &  4056.5 \\
extended basis, 59e-4c-CCSD(T)            & 0.09   &  11.1   \\
increased cutoff                          & 0.01   &  -0.5   \\
Gaunt                                    & -0.03  &  -0.5   \\
High order correlation effects           & -0.08  &  -2.7   \\
High Harmonics                           &  0.00  &  7.2    \\
\\
Total                                    & 4.50   & 4071.2  \\
\hline\hline
\end{tabular}
\end{table}
%
%
Taking into account the data from Table \ref{TResults} we estimate the uncertainty of the obtained value of $R_s$ at the level of 3\%. 

According to our estimations the correlation contribution to $R_s$ from 60 $1s..4f$ electrons of Fr is about 4\%. 
This is close to the contribution to the effective electric field acting on the electron EDM in the $^3\Delta_1$ electronic state of the ThO molecule from correlation of the $1s..4f$ electrons of Th found in Ref.~\cite{Skripnikov:16b}. Correlation contribution to $R_s$ from 28 $1s..3d$ electrons is about 2\%.

It was found that the correlation contributions to $R_s$ from the $4s..4f$ electrons are close within the AE4Z basis set and the CVTZ basis set (1.8\% vs. 1.4\%, correspondingly).
In particular, additional correlation functions for shell with the principal quantum number equal to 4 of Fr in the AE4Z basis set are not very important for the calculation of the $R_s$ value.  Besides, correlation contribution to $R_s$ from the $4s..4f$ electrons obtained at the CCS level of theory is almost equal to that obtained at the CCSD level within the strategy I. 
Thus, one can suggest that the most important effects from inclusion of the core electrons in the correlation treatment
%
are the spin polarization (orbital relaxation) effects that are described by the CCS method (in strategy I). This also justifies the use of the uncontracted CVTZ basis set to consider the correlation effects from the $1s..3d$ electrons: The basis set includes most of the important functions to describe the above mentioned  effects for $s$, $p$ (and $d$) waves which mainly contribute to $R_s$. A similar conclusion was made in Ref.~\cite{Skripnikov:16b}.
We have also found (within the CVDZ basis set \cite{Dyall:07,Dyall:12}) that the addition of the functions with high angular momenta for correlation of $1s..3d$ electrons from the AE4Z basis set to the CVDZ basis set 
contribute negligibly to $R_s$ though contribute significantly to the differential correlation energy of the electrons.

In Ref. \cite{Fleig:14} according to the orbital perturbation theory consideration it was concluded that the inner core correlation contribution to the effective electric field (analog of the $R_s$ constant) of ThO molecule is strongly suppressed. In that estimate the authors considered the admixture of the core orbital ($1s$) to the considered singly occupied orbital
(see \cite{Fleig:14} for details).
Below we show that one should consider another mechanism which describes the spin-polarization effect.
The mechanism can be formulated in the ``unified'' way which can be used to describe both molecules and atoms, though in the latter case one usually consider the third order perturbation theory, see e.g. Ref.\cite{Buhmann:02}.
Within strategy I the interaction with the external electric field $E_{\rm ext}$ is already included at the Dirac-Fock stage and after that one includes the $\hat{h}_{T,P}$ operator at the correlation stage.
From this point of view the present atomic problem is similar to the diatomic molecular one. In the latter case the spherical symmetry is also violated at the very beginning: $s$ and $p$ waves of the considered atom are mixed by the field of the second atom. In the present consideration ``zero-order'' orbitals are space-polarized orbitals as in the case of a polar molecule but spin unpolarized 
within the considered 87e-Kramers-restricted Dirac-Fock model \cite{DIRAC12}. In this picture the main contribution (from the valence electrons) to $R_s$ can be calculated as the expectation value of the $\hat{h}_{T,P}$ operator on the space-polarized orbital $\psi_0$ corresponding to the unpaired ``7$s$'' electron. Spin polarization of a (space-polarized) core orbital leads to a nonzero contribution to $R_s$ which can be estimated by the following perturbation theory expression:
\begin{eqnarray}
\nonumber 
 R_s(\rm core,PT)\cdot E_{\rm ext}=\\ 
 -
 \sum_{j>0} \frac{\langle \psi_{\rm core} | \hat{h}_{T,P} | \psi_{\rm virt,j} \rangle \langle \psi_0 \psi_{\rm virt,j} | \hat{V} | \psi_{\rm core} \psi_0 \rangle}{\epsilon^{(0)}_{\rm core} - \epsilon^{(0)}_{\rm virt,j}}+h.c.,
 \label{RsPT}
\end{eqnarray}
where $R_s(\rm core,PT)$ is the contribution to $R_s$ due to the spin polarization of the (space-polarized by the external field $E_{\rm ext}$) core orbital $\psi_{\rm core}$ with orbital energy $\epsilon^{(0)}_{\rm core}$, $\hat{V}=\frac{1}{|\textbf{r}_1-\textbf{r}_2|}$,
 $\psi_{\rm virt,j}$ is the (space-polarized) virtual orbital with the orbital energy $\epsilon^{(0)}_{\rm virt,j}$.

\begin{figure}
\includegraphics[scale=0.8]{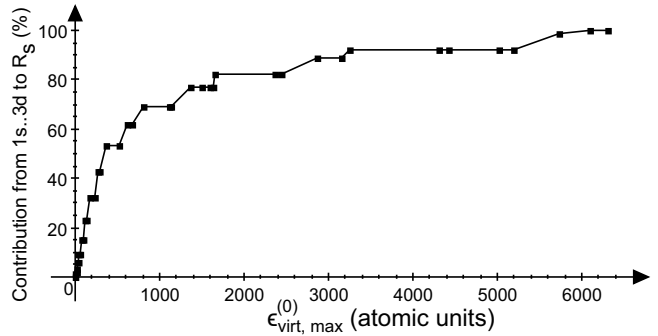}
\caption{The dependence of the contribution to $R_s$ from the spin polarization of the space-polarized $1s..3d$ orbitals of Fr on $\epsilon^{(0)}_{\rm virt,max}$, in percentage of the total PT2 contribution from the electrons.  $\epsilon^{(0)}_{\rm virt,max}$ is the energy of highest considered virtual orbital in Eq.~(\ref{RsPT}). 100\% corresponds to $\epsilon^{(0)}_{\rm virt,max}=6500$ a.u.
}
\label{RsCoreContrib} 
\end{figure}

The dependence of the contribution to $R_s$ from the spin polarization of the $1s..3d$ electrons of Fr 
calculated within Eq. (\ref{RsPT}) on the
 the orbital energy of the highest considered virtual orbital in Eq.~(\ref{RsPT})
 is given in Fig. \ref{RsCoreContrib}.
The most important contribution to $R_s(\rm core,PT)$ is from high energy virtual orbitals $\psi_{\rm virt,j}$.
Though their energies lead to big denominators the numerator can also be big.
For example, the matrix element $\langle \psi_{\rm core} | \hat{h}_{T,P} | \psi_{\rm virt,j} \rangle$ which includes $\psi_{\rm virt,j}$ with $\epsilon^{(0)}_{\rm virt,j} \approx 3000$  a.u. ($\epsilon^{(0)}_{\rm ``1s''} \approx -3742$ a.u.) is about seven orders greater than the main contribution given by the matrix element  $\langle\psi_{0} | \hat{h}_{T,P} | \psi_{0} \rangle$. 
Qualitatively, this is because only the virtual orbitals with high energies contribute to relaxation of the core orbitals in the region close to the heavy nucleus. As $\psi_0$ corresponds to the unpaired orbital with fixed spin projection, the relaxation effects for two paired core electrons are different and this leads to their spin polarization. 
Therefore, the core electrons \textit{can} lead to nonnegligiable contribution to $R_s$ (and they do; see above).

One should note that equations similar to Eq. (\ref{RsPT}) can also be used for estimation of the contribution to other spin-dependent AIC properties like the hyperfine structure constant, etc., in atoms and molecules.

Correlation effects for valence electrons should be considered at a significantly higher level of theory that can be done for the core electrons.
One can see from Table \ref{TResults} that the \textit{iterative triple} and noniterative quadruple cluster 
amplitudes contribute significantly (about 2\%) to the AIC characteristic $R_s$. Therefore, the high-order correlation effects for the valence electrons should be considered in accurate calculations
$R_s$ and 
other similar characteristics.

The final value is obtained to be $R_s = 4.50\cdot 10^{-18} e\cdot {\rm cm}$. 
Interestingly, it is close to the value of $4.5\cdot 10^{-18} e\cdot {\rm cm}$ obtained in Ref. \cite{Dzuba:11} 
(\cite{Note1}
) 
 by recalculation from the electron EDM enhancement factor 
from
Ref.~\cite{Byrnes:99} within the sum-over-states method.
Assuming that the scalar-pseudoscalar nucleus-electron interaction is the only source of Fr EDM,  $d_{\rm Fr}= k_{T,P} \cdot 4.50\cdot10^{-18} e\cdot {\rm cm}$. For example, assuming experimental limitation on $d_{\rm Fr} < 10^{-26} e\cdot {\rm cm}$ 
(e.g., present experimental limitation on 
the Hg atom EDM is $d_{\rm Hg} < 7.4\cdot 10^{-30} e\cdot {\rm cm}$ \cite{Graner:2016}) leads to the limitation on $k_{T,P} < 2.2 \cdot 10^{-9}$ which is smaller than the best limit obtained in the experiments on the ThO molecular beam ($k_{T,P} < 1.5 \cdot 10^{-8}$)~\cite{ACME:14a}, where we have recalculated corresponding limitation on the constant $C_S$ in~\cite{ACME:14a} to the constant $k_{T,P}$ defined by Eq.(\ref{Htp})).

\section{Conclusion}
We have expressed the measurable energy shift due to the T,P-odd interaction in the Fr atom in terms of the constant of the scalar-pseudoscalar interaction $k_{T,P}$. The combined four-component and two-step approach was applied and the two strategies of inclusion the interaction with the external electric field were considered for the calculation. 
Correlation contributions to the considered effect from groups of electrons were obtained. For core electrons the main contribution is due to their spin polarization.
Very good agreement with the available experimental data for the calculated ionization potential was obtained. 
A similar approach can be used to investigate other properties of Fr and related systems with high accuracy.

\section*{Acknowledgement}
We are grateful to Professor I. I. Tupitsyn for providing us the QED contribution to the ionization potential and Dr. A. N. Petrov for valuable discussions. Atomic calculations were partly performed on the Supercomputer ``Lomonosov.'' The 59-electron GRECP for Fr was generated in the framework of RFBR Grant No. 16-03-00766. The development of the code for the computation of the matrix elements of the considered operators as well as  the performance of all-electron calculations were funded by RFBR, according to research Project No.~16-32-60013 mol\_a\_dk. Two-step GRECP calculations were performed with the support of President of the Russian Federation Grant No.~MK-7631.2016.2 and Dmitry Zimin ``Dynasty'' Foundation.


\end{document}